\documentclass[10pt,final,doublecolumn]{IEEEtran}
\usepackage{amsmath}
\usepackage{latexsym}
\usepackage{graphicx}
\usepackage{bbding}
\usepackage{enumerate}
\usepackage{multicol}
\usepackage{color}
\usepackage{slashbox}
\usepackage{algorithm}
\usepackage{algorithmic}
\usepackage{subeqnarray}
\usepackage{cases}
\IEEEoverridecommandlockouts
\begin{document}

\title{Joint Optimization of Spectrum Sensing and Accessing in Multiuser MISO Cognitive Networks}

\author{\normalsize
Xiaoming~Chen, \emph{Senior Member, IEEE}, and Chau~Yuen,
\emph{Senior Member, IEEE}
\thanks{Xiaoming~Chen (e-mail: {\tt chenxiaoming@nuaa.edu.cn}) is
with the College of Electronic and Information Engineering, Nanjing
University of Aeronautics and Astronautics, and also with the
National Mobile Communications Research Laboratory, Southeast
University, China. Chau~Yuen (e-mail: {\tt yuenchau@sutd.edu.sg}) is
with Singapore University of Technology and Design,
Singapore.}}\maketitle

\begin{abstract}
In this paper, a joint spectrum sensing and accessing optimization
framework for a multiuser cognitive network is proposed to
significantly improve spectrum efficiency. For such a cognitive
network, there are two important and limited resources that should
be distributed in a comprehensive manner, namely feedback bits and
time duration. First, regarding the feedback bits, there are two
components: sensing component (used to convey various users' sensing
results) and accessing component (used to feedback channel state
information). A large sensing component can support more users to
perform cooperative sensing, which results in high sensing
precision. However, a large accessing component is preferred as
well, as it has a direct impact on the performance in the multiuser
cognitive network when multi-antenna technique, such as zero-forcing
beamforming (ZFBF), is utilized. Second, the tradeoff of sensing and
accessing duration in a transmission interval needs to be
determined, so that the sum transmission rate is optimized while
satisfying the interference constraint. In addition, the above two
resources are interrelated and inversive under some conditions.
Specifically, sensing time can be saved by utilizing more sensing
feedback bits for a given performance objective. Hence, the
resources should be allocation in a jointly manner. Based on the
joint optimization framework and the intrinsic relationship between
the two resources, we propose two joint resource allocation schemes
by maximizing the average sum transmission rate in a multiuser
multi-antenna cognitive network. Simulation results show that, by
adopting the joint resource allocation schemes, obvious performance
gain can be obtained over the traditional fixed strategies.

\end{abstract}


\section{Introduction}
Within the past decade, wireless communication has undergone an
unprecedentedly rapid growth to satisfy  users' demands for various
advanced services and applications. A stark reality resulted from
the explosive development of wireless communication is that there is
not enough available spectrum resource to ensure quality of services
required by the users' applications. However, FCC's report
\cite{FCC} indicates that the radio spectrum is heavily
under-utilized in time, frequency or space scales, which motivates a
spectrum open policy that allows unlicensed users to
opportunistically access the licensed spectrum when primary users
are inactive. Inspired by that idea, a novel communication
technology, namely cognitive radio, has been proposed and received
considerable research attentions from both academe and industry
\cite{Mitola}-\cite{Cogntivetrend}.

Since cognitive network is allowed to coexist with primary network
by the spectrum administrator, it is important for the cognitive
users to be transparent to the primary users by avoiding to create
interference that will degrade the performance of primary network.
In order to satisfy that rigorous requirement, cognitive networks
are designed to include two crucial components: spectrum sensing
\cite{sensing1}-\cite{sensing2} and dynamic spectrum accessing
\cite{accessing1}-\cite{accessing2}. Spectrum sensing has two
functions: 1) cognitive users should have strong enough ability to
detect active primary users to guarantee normal communication of
primary users, i.e. to increase the probability of detection; 2)
cognitive users need to try their best to find the unoccupied
spectrum so as to improve spectrum utilization efficiency, i.e. to
decrease the probability of false alarm. So far, there have been
several feasible spectrum sensing methods, for example, energy
detection \cite{energydetection1} \cite{energydetection2}, is a
simple and popular method, which has a preferable performance
without requiring any knowledge of primary signal a prior.
Unfortunately, it has a low sensing precision when the ratio of
sensing signal power to noise variance (sensing SNR) is low or noise
variance is uncertain \cite{lowSNR}. Although we can improve sensing
accuracy by increasing sensing duration, it will reduce the duration
for spectrum accessing for a given total duration constraint.

Recently, cognitive network equipped with multiple antennas is
proved to have the ability of further enhancing the performance of
both spectrum sensing \cite{MIMOSensing1} \cite{MIMOsensing2} and
accessing \cite{MIMOaccessing1} \cite{MIMOaccessing2} without adding
extra resource. By making use of its spatial dimensions, multiple
copies of primary signal are obtained and fused at the detector,
namely cooperative sensing
\cite{cooperativesensing1}-\cite{cooperativesensing2}. If these
copies are independent to each other, sensing time can be reduced
accordingly, for a given sensing precision. As a result, there is
more time available for accessing to improve transmission rate.
Especially, in the multiuser multiantenna paradigm, there are
several available performance-enhancing techniques for both spectrum
sensing and accessing. On one hand, multiuser cooperative sensing
can compensate for the insufficient capability of single user
through combining multiple sensing information. On the other hand,
multiuser MIMO technique is in favor of improving the transmission
rate during spectrum accessing \cite{multiuser1} \cite{multiuser2}.
It is worth pointing out that the performance of multiuser
multiantenna cognitive network is greatly dependent on feedback
resource \cite{CSIFeedback}, as both spectrum sensing and accessing
require the related information feedback. Specifically, if there are
more feedback amount available for spectrum sensing, more users are
allowed to convey their sensing results to cognitive base station
(BS), so that sensing precision can be improved. Meanwhile, if there
are more feedback amount available for spectrum sensing, users can
convey more accurate channel state information (CSI) to cognitive
BS. Cognitive BS performs to pre-process the signal to be
transmitted in order to decrease interuser interference, and thus
improve the performance. In order to optimize the performance, it is
necessary to allocate the feedback resource between spectrum sensing
and accessing for a given feedback resource constraint.

Common to most of the previous work on cognitive network is to
separately study spectrum sensing and accessing. In fact, the two
phases have a tight connection, especially for the multiuser
scenario. In order to achieve the optimal performance, it is
imperative to allocate the limited resource in the comprehensive
sense, namely joint optimization of spectrum sensing and accessing.
As discussed above, if there is a constraint on the total feedback
amount, we should determine the proportion of feedback amount
between sensing and accessing phases, so that the transmission rate
is maximized. Similarly, for each transmission interval, we also
need to determine the optimal interval of sensing and accessing
duration. As a result, the problem of joint resource allocation for
spectrum sensing and accessing is getting attention. As a pioneering
work, \cite{Jointoptimiztion1} initiated the problem of time
allocation between the two phases based on energy sensing to
maximize the throughput of cognitive user. \cite{Jointoptimization2}
extended the problem to the multiuser multichannel scenario. The
previous works in the literatures only consider one dimensional
resource allocation, for example the time dimension of spectrum
sensing or spectrum accessing. In fact, the allocation of feedback
amount is of equal importance to improve the performance of the
multiuser MISO cognitive network together with time. Considering the
correlation between the two resources, performance loss is
inevitable if the allocation of feedback amount and time duration is
optimized separately. In this paper, we address a joint spectrum
sensing and accessing problem in a multiuser MISO cognitive network.
By taking the maximization of the sum rate as the optimization
objective, we construct a joint optimization framework of spectrum
sensing and accessing, analyze the intrinsic relationship between
feedback and time resources, and then derive two joint resource
allocation schemes, which provide performance gain over the
traditional fixed resource allocation schemes.

The rest of this paper is organized as follows. Section II gives a
brief overview of the considered system model, and analyzes the
adopted spectrum sensing and accessing strategies. Then, by
maximizing the average sum transmission rate under some performance
requirements, we derive two joint resource allocation schemes in
Section III. Next, some simulation results are provided to verify
the effectiveness of the proposed schemes in Section IV and the
whole paper is concluded finally in Section V.

\textit{Notation}: We use bold upper (lower) letters to denote
matrices (column vectors), $(\cdot)^H$ to denote conjugate
transpose, $(\cdot)^{'}$ to denote the derivation, $E[\cdot]$ to
denote expectation, $\|\cdot\|$ to denote the $L_2$-norm of a
vector, and $|\cdot|$ to denote the absolute value. The acronym
i.i.d. means ``independent and identically distributed", pdf means
``probability density function" and cdf means ``cumulative
distribution function".

\section{System Model}
\begin{figure}[h] \centering
\includegraphics [width=0.45\textwidth] {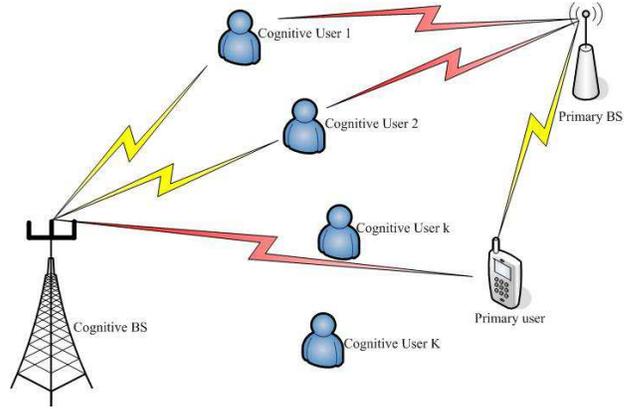}
\caption {The considered system block diagram.} \label{Fig1}
\end{figure}

In this paper, we consider a system including two networks, namely
primary and cognitive networks, as seen in Fig.\ref{Fig1}. $N_t$
antennas are mounted at the cognitive BS and $K$ cognitive users
equip with single antenna each. For ease of analysis, both primary
BS and primary user are considered to have single antenna. It is
assumed that the two networks are synchronized, and their
transmissions are both in the form of time slot of length $T$. At
the beginning of each time slot, multiple cognitive users
cooperatively sense the state of the licensed spectrum in the
duration of $\tau$. If the spectrum is regarded as busy, then the
cognitive network keeps silent in order to avoid the interference to
primary network. Otherwise, the cognitive network transmits in the
residual duration of $T-\tau$, so that the spectrum utilization
efficiency is improved effectively.

Note that in an interweave cognitive network, the accessing
opportunity of cognitive network is determined by the activity of
primary network, but not the traffic characteristics, so we only
preset the activity model \cite{activity}. First, the activity of
primary network is assumed to be fully independent of cognitive
network. Similar to \cite{activity}, the channel occupancy by
primary network is modeled as an ``alternating renewal source" that
alternates between busy and idle modes, where busy or idle denotes
the channel is occupied or not by primary network, respectively. We
use exponential distribution to describe the probability density
function (pdf) of the busy and idle periods of each time slot, which
can be expressed as
\begin{equation}
f_B(t)=\alpha{e}^{-\alpha t},\label{eqn1}
\end{equation}
and
\begin{equation}
f_I(t)=\beta{e}^{-\beta t},\label{eqn2}
\end{equation}
where $\alpha$ and $\beta$ are the transition rates from busy to
idle and from idle to busy, respectively. Then, the stationary
probabilities for the spectrum to be busy and idle can be written
as:
\begin{equation}
P_B=\frac{\beta}{\alpha+\beta},\label{eqn3}
\end{equation}
and
\begin{equation}
P_I=\frac{\alpha}{\alpha+\beta},\label{eqn4}
\end{equation}
respectively.

For an interweave cognitive network, it includes two important
stages, namely spectrum sensing and accessing. In what follows, we
introduce the spectrum sensing and accessing strategies adopted in
this paper, respectively.

\subsection{Spectrum Sensing}
Due to the simplicity, we consider the use of energy detection
\cite{energydetection1} \cite{energydetection2} for spectrum
sensing. In order to further improve the spectrum efficiency and
decrease the collision between the two networks, multiuser
cooperative sensing is adopted in the MISO cognitive network.

By making use of the Nyquist sampling technique, for cognitive user
$k$, the binary hypothesis test for spectrum sensing at time instant
$i$ takes the following form:
\begin{eqnarray}
\mathcal{H}_0&:&y_k(i)=n_k(i)\nonumber\\
\mathcal{H}_1&:&y_k(i)=g_k(i)s(i)+n_k(i),\label{eqn5}
\end{eqnarray}
where $y_k(i)$ is the received signal at cognitive user $k$, $s(i)$
is the transmit signal from primary BS, $g_k(i)$ is the channel gain
from primary BS to cognitive user $k$, and $n_k(i)$ is the zero mean
additive white Gaussian noise, i.e.,
$n_k(i)\sim\mathcal{CN}(0,\sigma_k^2)$, which is independent of
$s(i)$. Cognitive user $k$ obtains the statistics test $T_k$ by
summing the $N=2W\tau$ samplings, which is given by
\begin{equation}
T_k=\sum\limits_{i=0}^{N-1}|y_k(i)|^2,\label{eqn6}
\end{equation}
where $W$ is the spectrum bandwidth. If the number of sample is
large enough, according to the central limit theorem, the
distribution of $T_k$ can be approximated as
\begin{equation}
T_k\sim\left\{\begin{array}{ll}
                \mathcal{N}(N\sigma_k^2,N\sigma_k^4),& \mbox{$\mathcal{H}_0$}\\
                \mathcal{N}(N(\sigma_k^2+\sigma_s^2),N(\sigma_k^2+\sigma_s^2)^2), & \mbox{$\mathcal{H}_1$} \end{array}
                \right. \label{eqn7}
\end{equation}
where $\sigma_s^2$ is the variance of the sensing signal
$g_k(i)s(i)$. By employing energy detection, cognitive user $k$
judges the spectrum state is 1 (denote busy) if $T_k$ is greater
than a threshold $\lambda$, or 0 (denote idle). Sequently, the
sensing result is conveyed to cognitive BS by using 1 bit.

After receiving the feedback information from $L$ cognitive users,
the cognitive BS computes the final sensing result based on the
``or" fusion criteria \cite{Cooperativesensing}. Specifically, only
all $L$ cognitive users consider the spectrum is idle, the spectrum
state can be regarded as 0. Otherwise, the spectrum can not be
utilized by cognitive network. Through such a cooperative sensing
strategy, the detection probability $P(1|\mathcal{H}_1)$ and
false-alarm probability $P(1|\mathcal{H}_0)$ can be expressed as
\begin{equation}
P(1|\mathcal{H}_1)=1-\prod\limits_{l=1}^LP_l(0|\mathcal{H}_1),\nonumber
\end{equation}
and
\begin{equation}
P(1|\mathcal{H}_0)=1-\prod\limits_{l=1}^LP_l(0|\mathcal{H}_0),\nonumber
\end{equation}
where $P_l(0|\mathcal{H}_1)$ and $P_l(0|\mathcal{H}_0)$ are the
probabilities that cognitive user $l$ judges the spectrum is idle
when primary network is active and inactive, respectively. It is
reasonably assumed that the cognitive users have the same
$P_l(0|\mathcal{H}_1)$ and $P_l(0|\mathcal{H}_0)$ in the statistical
sense due to their similar sense capabilities. Thereby, the
detection and false-alarm probabilities can be rewritten as
\begin{eqnarray}
P(1|\mathcal{H}_1)&=&1-(1-P_d)^L\nonumber\\
&=&1-\left(1-Q\left(\frac{\lambda-N(\sigma_n^2+\sigma_s^2)}{\sqrt{N(\sigma_s^2+\sigma_n^2)^2}}\right)\right)^L,\label{eqn8}
\end{eqnarray}
and
\begin{eqnarray}
P(1|\mathcal{H}_0)&=&1-(1-P_f)^L\nonumber\\
&=&1-\left(1-Q\left(\frac{\lambda-N\sigma_n^2}{\sqrt{N\sigma_n^4}}\right)\right)^L\nonumber\\
&\approx&LQ\left(\frac{\lambda-N\sigma_n^2}{\sqrt{N\sigma_n^4}}\right),\label{eqn9}
\end{eqnarray}
where
$P_d=Q\left(\frac{\lambda-N(\sigma_n^2+\sigma_s^2)}{\sqrt{2N(\sigma_s^2+\sigma_n^2)^2}}\right)$
and
$P_f=Q\left(\frac{\lambda-N\sigma_n^2}{\sqrt{2N\sigma_n^4}}\right)$
are respectively the detection and false-alarm probabilities of an
arbitrary cognitive user, which are derived based on the
distribution of $T_k$ in (\ref{eqn7}) and the assumption that all
cognitive users have the same variance of the receive noise
$\sigma_n^2$, and
$Q(x)=\frac{1}{\sqrt{2\pi}}\int_x^\infty\exp(-\frac{y^2}{2})dy$ is
the Q-function. (\ref{eqn9}) follows from the fact $(1-x)^L\approx
1-Lx$ when $x$ is a sufficient small value. Based on (\ref{eqn8})
and (\ref{eqn9}), we have the following relationship
\begin{eqnarray}
P(1|\mathcal{H}_0)&=&
LQ\bigg\{\frac{Q^{-1}\left(1-\left(1-P(1|\mathcal{H}_1)\right)^{1/L})\right)}{\sqrt{N\sigma_n^4}}\nonumber\\
&&\times\frac{\sqrt{N(\sigma_s^2+\sigma_n^2)^2}+N\sigma_s^2}{\sqrt{N\sigma_n^4}}\bigg\}\nonumber\\
&=&LQ\bigg\{Q^{-1}\left(1-\left(1-P(1|\mathcal{H}_1)\right)^{1/L}\right)\nonumber\\
&&\times\left(1+\xi\right)+\sqrt{2W\tau}\xi\bigg\},\label{eqn10}
\end{eqnarray}
where $\xi=\frac{\sigma_s^2}{\sigma_n^2}$ is the so-called sensing
SNR of the received sensing signal.

\subsection{Spectrum Accessing}
If primary network is judged as inactive based on the fused sensing
information, cognitive network accesses the spectrum
opportunistically. For a multiuser network, it is proved that the
system performance is improved with the increase of feedback amount
about channel state information. In this paper, we adopt limited
feedback zero-forcing beamforming (ZFBF) as the spectrum accessing
strategy. For convenience of analysis, we assume that all cognitive
users always have information to receive and they are scheduled
based on round robin policy. Specifically, cognitive users are
serviced by an predetermined order independent of channel state.
During each time slot, a fixed number of cognitive users, such as
$N_t$, are coordinated to access the available spectrum based on a
certain user scheduling scheme, e.g. round bin.

In this paper, we adopt codebook based limited feedback scheme. When
the spectrum is open, cognitive user $k$, who is scheduled in
current slot, selects an optimal codeword
$\hat{\textbf{h}}_{k,\textrm{opt}}$ from the predetermined codebook
$\mathcal{H}_k$ of size $2^B$ based on the instantaneous CSI
$\textbf{h}_k$, where
$\mathcal{H}_k=\{\hat{\textbf{h}}_{k,1},\hat{\textbf{h}}_{k,2},\cdots,
\hat{\textbf{h}}_{k,2^B}\}$. The codeword selection criteria can be
expressed as
\begin{equation}
\hat{\textbf{h}}_{k,\textrm{opt}}=\arg\max\limits_{1\leq j\leq
2^B}|\tilde{\textbf{h}}_k^H\hat{\textbf{h}}_{k,j}|^2,\label{eqn30}
\end{equation}
where $\tilde{\textbf{h}}_k=\frac{\textbf{h}_k}{\|\textbf{h}_k\|}$
is the direction vector of $\textbf{h}_k$. The other scheduled users
select their optimal codewords from different quantization codebooks
and convey the corresponding selected codeword indexes to cognitive
BS. Assuming that the channels are i.i.d. and block fading. In other
words, the channel keeps constant during a time slot and fades
independently slot by slot. It is worth pointing out that the
duration of CSI feedback is quite small and is negligible, we do not
consider it in this paper. After receiving the feedback information
of $N_t$ current scheduled cognitive users, cognitive BS determines
the optimal transmit beams $\textbf{w}_{k}, k=1,\cdots, N_t$ by
making use of ZFBF design method. Specifically, for the $k$th user,
we first construct its complementary channel matrix
\begin{equation}
\bar{\hat{\textbf{H}}}_k=[\hat{\textbf{h}}_{1,\textrm{opt}}, \cdots,
\hat{\textbf{h}}_{k-1,\textrm{opt}},
\hat{\textbf{h}}_{k+1,\textrm{opt}}, \cdots,
\hat{\textbf{h}}_{N_t,\textrm{opt}}],\nonumber
\end{equation}
where $\hat{\textbf{h}}_{k-1,\textrm{opt}}$ is the optimal codeword
selected by the $(k-1)$th user based on (\ref{eqn30}). Taking
singular value decomposition (SVD) to $\bar{\hat{\textbf{H}}}_k$, if
$\textbf{V}_{k}^{\perp}$ is the matrix composed of the right
singular vectors with respect to zero singular values, then
$\textbf{w}_k$ is a normalized vector spanned by the space of
$\textbf{V}_{k}^{\perp}$, so that we have
\begin{equation}
\hat{\textbf{h}}_{u,\textrm{opt}}^H\textbf{w}_k=0,\quad k\leq u\leq
N_t, u\neq k\nonumber
\end{equation}
It is assumed that $x_k$ is the expected normalized signal of the
$k$th current scheduled user, then its received signal can be
expressed as
\begin{equation}
y_k=\sqrt{\frac{P}{N_t}}\sum\limits_{u=1}^{N_t}\textbf{h}_k^H\textbf{w}_ux_u+n_k+n_s,\label{eqn11}
\end{equation}
where $P$ is the total transmit power of cognitive BS, which is
equally allocated to $N_t$ users. $n_k$ is the additive Gaussian
white noise with zero mean and covariance $\sigma_n$ for all users.
Due to miss-detection, we consider the interference $n_s$ from
primary network. For ease of analysis, it is assumed that $n_s$ is
an i.i.d. complex Gaussian random variable with zero mean and
covariance $\sigma_s$. Hence, the ratio of the received signal to
interference and noise (SINR) for the $k$th user can be expressed as
\begin{eqnarray}
\rho_k&=&\frac{|\textbf{h}_k^H\textbf{w}_k|^2}{N_t(\sigma_s^2+\sigma_n^2)/P+\sum\limits_{u=1,u\neq k}^{N_t}|\textbf{h}_k^H\textbf{w}_u|^2}\nonumber\\
&=&\frac{|\textbf{h}_k^H\textbf{w}_k|^2}{1/\gamma+\sum\limits_{u=1,u\neq
k}^{N_t}|\textbf{h}_k^H\textbf{w}_u|^2},\label{eqn12}
\end{eqnarray}
where
$\gamma=\frac{P}{N_t(\sigma_n^2+\sigma_s^2)}=\frac{\varrho}{N_t(1+\xi)}$,
$\varrho=\frac{P}{\sigma_n^2}$ is the average transmit SNR at
cognitive BS, and $\xi=\frac{\sigma_s^2}{\sigma_n^2}$ is the sensing
SNR.

As a result, the average sum transmission rate of the multiuser MISO
cognitive network based on limited feedback ZFBF in the presence of
mis-detection can be expressed as
\begin{eqnarray}
R&=&E\left[\sum\limits_{k=1}^{N_t}\log_2\left(1+\rho_k\right)\right]\nonumber\\
&=&N_tE\left[\log_2(1+\rho_k)\right],\label{eqn13}
\end{eqnarray}
where (\ref{eqn13}) holds true because all the channels are i.i.d.
Clearly, in order to compute the average sum transmission rate, the
key is the achievement of the pdf of the received SINR $\rho_k$. For
the pdf and cdf of $\rho_k$, we have the following lemma:

\emph{Lemma 1}: For limited feedback zero-forcing beamforming in the
setting of $N_t$ BS antennas, $N_t$ single antenna users and the
quantization codebooks of size $2^B$, the pdf and cdf of SINR are
$f_{\rho_k}(x)=1/\gamma\exp(-x/\gamma)(1+\delta
x)^{-(N_t-1)}+\delta(N_t-1)\exp(-x/\gamma)(1+\delta x)^{-N_t}$ and
$F_{\rho_k}(x)=1-\frac{\exp(-x/\gamma)}{(1+\delta x)^{N_t-1}}$
respectively, where $\delta=2^{-\frac{B}{N_t-1}}$.

\begin{proof}
Please refer to Appendix.
\end{proof}

Based on the above pdf and cdf of the received SINR, the average sum
transmission rate can be computed as
\begin{eqnarray}
R&=&N_tE\left[\log_2(1+\rho_k)\right]\nonumber\\
&\approx&N_t\log_2(e)E\left[\ln(1+\rho_k)\right]\nonumber\\
&=&N_t\log_2(e)\int_0^{\infty}\ln(1+x)f_{\rho_k}(x)dx\nonumber\\
&=&-N_t\log_2(e)\int_0^{\infty}\ln(1+x)\left(1-F_{\rho_k}(x)\right)^{'}dx\nonumber\\
&=&N_t\log_2(e)\int_0^{\infty}\frac{1-F_{\rho_k}(x)}{1+x}dx\nonumber\\
&=&\frac{N_t\log_2(e)}{\delta^{N_t-1}}\int_0^{\infty}\frac{\exp(-x/\gamma)}{(x+1)(x+
\delta^{-1})^{N_t-1}}dx\nonumber\\
&=&\frac{N_t\log_2(e)}{\delta^{N_t-1}}I_1\left(\gamma^{-1},\delta^{-1},N_t-1\right)\label{eqn14}\\
&=&2^BN_t\log_2(e)I_1\left(\gamma^{-1},2^{\frac{B}{N_t-1}},N_t-1\right),\label{eqn15}
\end{eqnarray}
where
\begin{eqnarray}
I_1(a,b,M)&=&\int_0^{\infty}\frac{\exp(-ax)}{(x+1)(x+b)^{M}}dx\nonumber\\
&=&\sum\limits_{i=1}^M(-1)^{i-1}(1-b)^{-i}I_2(a,b,M-i+1)\nonumber\\
&&+(b-1)^{-M}I_2(a,1,1),\label{eqn16}
\end{eqnarray}
\begin{equation}
I_2(a,b,M)=\left\{\begin{array}{ll} \exp(ab)E_1(ab) & M=1\\
\sum\limits_{i=1}^{M-1}\frac{(i-1)!}{(M-1)!}\frac{(-a)^{M-i-1}}{b^i}\\+\frac{(-a)^{M-1}}{(M-1)!}\exp(ab)\textmd{E}_1(ab),
& M\geq2
\end{array}\right.\label{eqn17}
\end{equation}
and $\textmd{E}_1(x)$ is the exponential-integral function of the
first order. Hence, we derive the average sum transmission rate in
presence of mis-detection as a function of codebook size $B$.

\section{Joint Optimization of Spectrum Sensing and Accessing}
This section concentrates on the parameter optimization of spectrum
sensing and accessing, so as to maximize the average sum
transmission rate. As analyzed above, spectrum sensing and accessing
have a tight connection and together they determine the system
performance. In order to maximize the average transmission rate, it
is imperative to jointly optimize spectrum sensing and accessing
through feedback amount and time duration allocation.

\subsection{The Fixed Number of Accessing Users}
\begin{figure}[h] \centering
\includegraphics [width=0.45\textwidth] {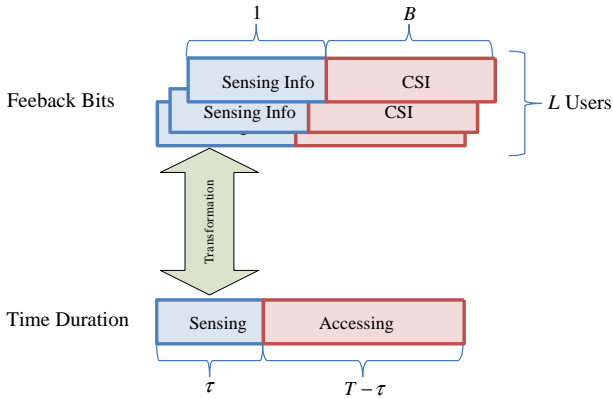}
\caption {The block diagram of joint resource allocation.}
\label{Fig4}
\end{figure}

For multiuser cooperative spectrum sensing, if one cognitive user
uses 1 bit to inform cognitive BS its sensing result, when there are
$L$ users involved in, the total feedback bits for spectrum sensing
will be $L$ bits for each time slot. For spectrum accessing based on
ZFBF, one scheduled user uses $B$ bits to quantize its instantaneous
CSI. Thereby, the total feedback amount for spectrum sensing and
accessing is $L+N_tB\leq\iota$, where $\iota$ is the constraint on
feedback amount. In addition, for time resource, we assume the
duration of sensing time is $\tau$ and the left time $T-\tau$ is
allocated for accessing, as seen in Fig.\ref{Fig4}. In this paper,
we maximize the average sum transmission rate while satisfying the
protection to primary network as the optimization objective, which
can be formulated as the following problem:
\begin{eqnarray}
J_1&=&\max\limits_{\tau, L, B}
\left(1-\frac{\tau}{T}\right)\bigg(P_I\left(1-P(1|\mathcal{H}_0)\right)R_1\nonumber\\
&&\quad\quad\quad+P_B\left(1-P(1|\mathcal{H}_1)\right)R_2\bigg)\nonumber
\end{eqnarray}
Subject to
\begin{subequations}
\begin{numcases}{}
P(1|\mathcal{H}_0)\leq P_0\label{eqn18}\\
P(1|\mathcal{H}_1)\geq P_1\label{eqn19}\\
0\leq\tau\leq T\label{eqn20}\\
L+N_tB\leq\iota,\label{eqn21}
\end{numcases}
\end{subequations}
where $P_0$ and $P_1$ are respectively the upper bound of
false-alarm probability and the lower bound of detection probability
preset to improve spectrum efficiency and protect the normal
communication of primary network.
$R_1=2^BN_t\log_2(e)I_1\left(N_t\sigma_n^2/P,2^{\frac{B}{N_t-1}},N_t-1\right)$
and
$R_2=2^BN_t\log_2(e)I_1\left(N_t(\sigma_n^2+\sigma_s^2)/P,2^{\frac{B}{N_t-1}},N_t-1\right)$
are the average sum transmission rates of cognitive network when
primary network is idle and busy, respectively, according to
(\ref{eqn15}). Examining the objective function, we find that it is
a decreasing function of detection function $P(1|\mathcal{H}_1)$,
and considering it is lower bounded by $P_1$, so the objective
function is maximized when $P(1|\mathcal{H}_1)=P_1$. Then, combining
(\ref{eqn10}) and (\ref{eqn18}), we have
\begin{equation}
LQ\left(Q^{-1}\left(1-\left(1-P(1|\mathcal{H}_1)\right)^{1/L}\right)\left(1+\xi\right)+\sqrt{2W\tau}\xi\right)\leq
P_0.\nonumber
\end{equation}
Solving the above inequality, we get
\begin{equation}
\tau\geq\frac{\left(Q^{-1}\left(P_0/L\right)-Q^{-1}\left(1-\left(1-P_1\right)^{1/L}\right)(1+\xi)\right)^2}{2W\xi^2}.\label{eqn22}
\end{equation}
Thus, the optimization problem is reduced as
\begin{eqnarray}
J_2&=&\max\limits_{\tau, L, B}
\left(1-\frac{\tau}{T}\right)\bigg(P_I\left(1-P(1|\mathcal{H}_0)\right)R_1\nonumber\\
&&\quad\quad\quad+P_B\left(1-P_1\right)R_2\bigg)\nonumber
\end{eqnarray}
Subject to
\begin{subequations}
\begin{numcases}{}
\frac{\bigg(Q^{-1}\left(P_0/L\right)-Q^{-1}\left(1-(1-P_1)^{1/L}\right)(1+\xi)\bigg)^2}{2W\xi^2}\nonumber\\
\leq\tau\leq T\label{eqn23}\\
L+N_tB\leq\iota.\label{eqn24}
\end{numcases}
\end{subequations}

Unfortunately, the above optimization problem is a mixed integer
programming problem, it is difficult to obtain a close-form
expression of optimal $\tau$, $L$ and $B$. Intuitively, the optimal
solution can be obtained by exhaustive search. Specifically, for a
given feedback combination, we can get the optimal sensing duration
by solving a convex optimization related to $\tau$. Then, through
comparing the average sum rate of all feedback combinations, we can
get the optimal one. However, if the feedback amount constraint
$\iota$ is large, the computation complexity is unbearable.
Alternatively, we attempt to seek a suboptimal method to jointly
decide the three parameters of spectrum sensing and accessing, so
that we can achieve a feasible solution for practical
implementation.

First, examining the objective function, it has the following
appealing property:

\emph{Lemma 2}: Given $L$ and $B$, $J_2$ is a concave function with
respective to $\tau$.

\begin{proof}
Replacing $P(1|\mathcal{H}_1)$ with
$LQ\left(Q^{-1}\left(1-\left(1-P_1\right)^{1/L}\right)\left(1+\xi\right)+\sqrt{2W\tau}\xi\right)$
according to (\ref{eqn10}), the objective function can be rewritten
as
$V(\tau)=\left(1-\frac{\tau}{T}\right)\bigg((1-LQ(Q^{-1}(1-(1-P_1)^{1/L})(1+\xi)+\sqrt{2W\tau}\xi))A+C\bigg)$,
where $A=P_IR_1$ and $C=P_B\left(1-P_1\right)R_2$ are two positive
constants independent of $\tau$. Taking two-order derivation to
$V(\tau)$ with respect to $\tau$, we have
\begin{eqnarray}
\frac{d^2V(\tau)}{d\tau^2}&=&-\left(1-\frac{\tau}{T}\right)\bigg\{\frac{L}{2\pi}\exp\left(-\frac{(U+D\sqrt{\tau})^2}{2}\right)\nonumber\\
&&\times\left(\frac{D^2}{2}+\frac{UD}{4}\tau^{-1/2}+\frac{\tau^{-1}}{2}\right)\frac{D}{2}\tau^{-1/2}\bigg\}\nonumber\\
&&-\frac{2}{T}\frac{L}{2\pi}\exp\left(-\frac{(U+D\sqrt{\tau})^2}{2}\right)\frac{D}{2}\tau^{-1/2}\nonumber\\
&<&0,
\end{eqnarray}
where $U=Q^{-1}\left(1-(1-P_1)^{1/L}\right)(1+\xi)$ and
$D=\sqrt{2W}\xi$. Hereby, $J_2$ is a concave function with respect
to $\tau$.
\end{proof}

Since the constraint (\ref{eqn23}) is linear, when given $L$ and
$B$, we can derive the optimal sensing time $\tau$ by Lagrange
multiplier method. However, given $\tau$, solving $L$ and $B$ is an
integer programming problem. Intuitively, $J_2$ is an increasing
function of $L$ and $B$ because they are beneficial to improve the
system performance, so the total feedback amount $\iota$ should be
utilized as completely as possible. To solve such an integer
programming problem, greedy algorithm is a simple and powerful
choice. Specifically, from a given initial values, at each step, $L$
or $B$ is added by 1 to compare the performance gain. If the
performance gain caused by 1 increment on $L$ is larger, then
$L=L+1$. Otherwise, $B=B+1$. Therefore, with the purpose of joint
optimization of $\tau$, $L$ and $B$, we can first allocate the
feedback bits by greedy algorithm for a given sensing time, and then
update the optimal sensing time based on the predetermined $L$ and
$B$. The iteration stops until all feedback bits are used. Thus, the
whole procedure can be summarized as below.\\
\rule{8.9cm}{1pt} \emph{Algorithm 1}\\\rule{8.9cm}{1pt}
\begin{enumerate}
\item Initialization: given $\alpha$, $\beta$, $W$, $P_0$, $P_1$,
$T$, $\varrho$, $\xi$ and $\iota$, and set $L=1$, $B=1$,
$\tau_l(L,B)=\frac{\left(Q^{-1}(P_0/L)-Q^{-1}(1-(1-P_1)^{1/L})(1+\xi)\right)^2}{2W\xi^2}$,
$\tau_u=T$ and $V(\tau, L,
B)=\left(1-\frac{\tau}{T}\right)\bigg((1-LQ(Q^{-1}(1-(1-P_1)^{1/L})(1+\xi)+\sqrt{2W\tau}\xi))
P_IR_1+P_B\left(1-P_1\right)R_2\bigg)$.

\item Let $L_0=L+1$, $B_0=B+1$, $\tau_{0,0}=\tau_l(L_0,B)$,
$\tau_{0,1}=\tau_u$, $\tau_{1,0}=\tau_l(L,B_0)$ and
$\tau_{1,1}=\tau_u$.

\item Compute $V(\tau_{0,0},L_0,B)$ and
$V(\tau_{0,1},L_0,B)$. If $V(\tau_{0,0},L_0,B)\leq
V(\tau_{0,1},L_0,B)$, then
$\tau_{0,0}=\frac{\tau_{0,0}+\tau_{0,1}}{2}$. Otherwise,
$\tau_{0,1}=\frac{\tau_{0,0}+\tau_{0,1}}{2}$. If
$\tau_{0,1}-\tau_{0,0}>\varepsilon$ ($\varepsilon$ is a quite small
real value), then repeat from 3), otherwise let
$\tau_0=\frac{\tau_{0,0}+\tau_{0,1}}{2}$.

\item Compute $V(\tau_{1,0},L,B_0)$ and
$V(\tau_{1,1},L,B_0)$. If $V(\tau_{1,0},L,B_0)\leq
V(\tau_{1,1},L,B_0)$, then
$\tau_{1,0}=\frac{\tau_{1,0}+\tau_{1,1}}{2}$. Otherwise,
$\tau_{1,1}=\frac{\tau_{1,0}+\tau_{1,1}}{2}$. If
$\tau_{1,1}-\tau_{1,0}>\varepsilon$ ($\varepsilon$ is a quite small
real value), then repeat from 4), otherwise let
$\tau_{1}=\frac{\tau_{1,0}+\tau_{1,1}}{2}$.

\item If $V(\tau_{0},L_0,B)\geq V(\tau_{1},L,B_0)$, then $L=L_0$ and
$\tau=\tau_{0}$. Otherwise, $B=B_0$ and $\tau=\tau_1$.

\item If $(L+1)+N_tB\leq\iota$ and $L+N_t(B+1)\leq\iota$, then repeat from 2).
If $(L+1)+N_tB\leq\iota$ and $L+N_t(B+1)>\iota$, then
$L=L_0=\iota-N_tB$, compute $\tau_0$ according to 3), let
$\tau=\tau_0$.

\end{enumerate}
\rule{8.9cm}{1pt}

\emph{Remark}: During the above joint optimization, we compute the
number of sensing users $L$ and codebook size $B$ by the greedy
algorithm. Note that the sensing duration is obtained by the
bisection method, so the computation amount of Algorithm 1 is
$\iota$ times as many as that of the bisection method at most.
Although it is not optimal, it achieves a preferable tradeoff
between system performance and implementation complexity to some
extent compared with the exhaustion method.

\subsection{The Variable Number of Accessing Users}
It is worth noting that in the above, we fix the number of accessing
users as $N_t$. In fact, $N_t$ just is the upper bound of the number
of accessing users $K_0$, which is a variable scaling from $1$ to
$N_t$ depending on the network condition. Generally speaking, if the
network is noise and interference (from primary network) limited, a
large number of accessing users is better to improve the
performance. Otherwise, if the network is interuser interference
limited, a small number of accessing users is preferable. The number
of accessing users has a direct impact on the allocation of feedback
bits, and thus time duration. In addition, the constraints on
detection and false-alarm probabilities would also affect the number
of accessing users. Hence, as a parameter of joint optimization of
spectrum sensing and accessing, it is imperative to determine the
optimal number of accessing user $K_0$ according to network
conditions and sensing constraints. If we assume the number of
accessing users is $K_0$, then the received SINR for the $k$th user
with miss-detection can be expressed as
\begin{equation}
\rho_k=\gamma|\textbf{h}_k^H\textbf{w}_k|^2,\label{eqn25}
\end{equation}
for $K_0=1$, where $\textbf{w}_k=\hat{\textbf{h}}_k$ is the maximum
ratio transmission (MRT). MRT is selected because it can maximize
the average transmission rate with limited feedback when no
inter-user interference cancelation is needed, and
\begin{eqnarray}
\rho_k&=&\frac{|\textbf{h}_k^H\textbf{w}_k|^2}{M(\sigma_s^2+\sigma_n^2)/P+\sum\limits_{u=1,u\neq k}^{K_0}|\textbf{h}_k^H\textbf{w}_u|^2}\nonumber\\
&=&\frac{|\textbf{h}_k^H\textbf{w}_k|^2}{1/\gamma+\sum\limits_{u=1,u\neq
k}^{K_0}|\textbf{h}_k^H\textbf{w}_u|^2},\label{eqn26}
\end{eqnarray}
for $2\leq K_0\leq N_t$.

For the case of $2\leq K_0\leq N_t$, based on the similar analysis
in Appendix, we could get the corresponding pdf and cdf of received
SINR in the case of $K_0$ accessing users. Averaging the
instantaneous sum transmission rate over the pdf of received SINR,
we have
\begin{equation}
R=2^{\frac{B(K_0-1)}{N_t-1}}K_0\log_2(e)I_1\left(\gamma^{-1},2^{\frac{B}{N_t-1}},K_0-1\right).\label{eqn26}
\end{equation}
Following \cite{LFBF}, the average transmission rate of MRT is given
by
\begin{eqnarray}
R&=&\log_2(e)\bigg\{\exp\left(\gamma^{-1}\right)\sum\limits_{k=0}^{N_t-1}E_{k+1}\left(\gamma^{-1}\right)\nonumber\\
&&-\int_0^1\left(1-(1-\upsilon)^{N_t-1}\right)^{2^B}\frac{N_t}{\upsilon}
\exp\left((\gamma\upsilon)^{-1}\right)\nonumber\\
&&\times
E_{N_t+1}\left((\gamma\upsilon)^{-1}\right)d\upsilon\bigg\},\label{eqn27}
\end{eqnarray}
where $E_n(x)=\int_1^{\infty}\exp(-xt)x^{-n}dt$ is the $n$th
exponential integral. In order to encourage the users to sense the
spectrum, we stipulate that only the sensing users are allowed to
access the available spectrum. In other words, the number of sensing
users $L$ is equal to the number of accessing users $K_0$. Hence,
the joint optimization of spectrum sensing and accessing is
equivalent to the following optimization problem
\begin{eqnarray}
J_3&=&\max\limits_{\tau, L, B, K_0}
\left(1-\frac{\tau}{T}\right)\bigg(P_I\left(1-P(1|\mathcal{H}_0)\right)R_1\nonumber\\
&&\quad\quad\quad+P_B\left(1-P_1\right)R_2\bigg)\nonumber
\end{eqnarray}
Subject to
\begin{subequations}
\begin{numcases}{}
\frac{\left(Q^{-1}\left(P_0/L\right)-Q^{-1}\left(1-(1-P_1)^{1/L}\right)(1+\xi)\right)^2}{2W\xi^2}\nonumber\\
\leq\tau\leq T\label{eqn28}\\
L+LB\leq\iota,\label{eqn29}
\end{numcases}
\end{subequations}
where
$R_1=2^{\frac{B(L-1)}{N_t-1}}L\log_2(e)I_1\big(L\sigma_n^2/P,2^{\frac{B}{N_t-1}},L-1\big)$
and
$R_2=2^{\frac{B(L-1)}{N_t-1}}L\log_2(e)I_1\big(L(\sigma_n^2+\sigma_s^2)/P,2^{\frac{B}{N_t-1}},L-1\big)$
for $2\leq K_0=L\leq N_t$, and
$R_1=\log_2(e)(\exp\left(\sigma_n^2/P\right)\sum\limits_{k=0}^{N_t-1}E_{k+1}\left(\sigma_n^2/P\right)
-\int_0^1\left(1-(1-\upsilon)^{N_t-1}\right)^{2^B}\frac{N_t}{\upsilon}
\exp\left(\sigma_n^2/(P\upsilon)\right)E_{N_t+1}\left(\sigma_n^2/(P\upsilon)\right)\\d\upsilon)$
and
$R_2=\log_2(e)(\exp((\sigma_n^2+\sigma_s^2)/P)\sum\limits_{k=0}^{N_t-1}E_{k+1}(\gamma^{-1}(\sigma_n^2+\sigma_s^2)/P)
-\int_0^1(1-(1-\upsilon)^{N_t-1})^{2^B}\frac{N_t}{\upsilon}
\exp((\sigma_n^2+\sigma_s^2)/(P\upsilon))E_{N_t+1}((\sigma_n^2+\sigma_s^2)/(P\upsilon))d\upsilon)$
for $K_0=L=1$. Similar to $J_2$, this problem is also a mixed
integer programming problem, so that it is difficult to give a
close-form expression of the optimal solution. Based on the same
idea of algorithm 1, the number of accessing users, the number of
sensing users, codebook size and sensing time can be jointly
determined through the following algorithm\\
\rule{8.9cm}{1pt} \emph{Algorithm 2}\\\rule{8.9cm}{1pt}
\begin{enumerate}
\item Initialization: given $\alpha$, $\beta$, $W$, $P_0$, $P_1$,
$T$, $\varrho$, $\xi$ and $\iota$, and set $L=1$, $B=1$,
$\tau_l(L,B)=\frac{\left(Q^{-1}(P_0/L)-Q^{-1}(1-(1-P_1)^{1/L})(1+\xi)\right)^2}{2W\xi^2}$,
$\tau_u=T$ and $V(\tau, L,
B)=\left(1-\frac{\tau}{T}\right)\bigg((1-LQ(Q^{-1}(1-(1-P_1)^{1/L})(1+\xi)+\sqrt{2W\tau}\xi))
P_IR_1+P_B\left(1-P_1\right)R_2\bigg)$.

\item Let $L_0=L+1$, $B_0=B+1$, $\tau_{0,0}=\tau_l(L_0,B)$,
$\tau_{0,1}=\tau_u$, $\tau_{1,0}=\tau_l(L,B_0)$ and
$\tau_{1,1}=\tau_u$.

\item Compute $V(\tau_{0,0},L_0,B)$ and
$V(\tau_{0,1},L_0,B)$. If $V(\tau_{0,0},L_0,B) \leq
V(\tau_{0,1},L_0,B)$, then
$\tau_{0,0}=\frac{\tau_{0,0}+\tau_{0,1}}{2}$. Otherwise,
$\tau_{0,1}=\frac{\tau_{0,0}+\tau_{0,1}}{2}$. If
$\tau_{0,1}-\tau_{0,0}>\varepsilon$ ($\varepsilon$ is a quite small
real value), then repeat from 3), otherwise let
$\tau_0=\frac{\tau_{0,0}+\tau_{0,1}}{2}$.

\item Compute $V(\tau_{1,0},L,B_0)$ and
$V(\tau_{1,1},L,B_0)$. If $V(\tau_{1,0},L,B_0)\leq
V(\tau_{1,1},L,B_0)$, then
$\tau_{1,0}=\frac{\tau_{1,0}+\tau_{1,1}}{2}$. Otherwise,
$\tau_{1,1}=\frac{\tau_{1,0}+\tau_{1,1}}{2}$. If
$\tau_{1,1}-\tau_{1,0}>\varepsilon$ ($\varepsilon$ is a quite small
real value), then repeat from 4), otherwise let
$\tau_{1}=\frac{\tau_{1,0}+\tau_{1,1}}{2}$.

\item If $V(\tau_{0},L_0,B)\geq V(\tau_{1},L,B_0)$, then $L=L_0$ and
$\tau=\tau_{0}$. Otherwise, $B=B_0$ and $\tau=\tau_1$.

\item If $L+1+(L+1)B\leq\iota$, $L+L(B+1)\leq\iota$ and $L+1\leq N_t$,
then repeat from 2). If $L+1+(L+1)B\leq\iota$, $L+L(B+1)>\iota$ and
$L+1\leq N_t$, then $L=L_0=\left\lfloor\frac{\iota}{B+1}\right
\rfloor$, compute $\tau_0$ according to 3), let $\tau=\tau_0$. If
$L+1+(L+1)B>\iota$ and $L+L(B+1)\leq\iota$, then
$B=B_0=\left\lfloor\frac{\iota}{L}\right\rfloor$-1, compute $\tau_1$
according to 4), let $\tau=\tau_1$.

\item Set $K_0=L$.

\end{enumerate}
\rule{8.9cm}{1pt}

Notice that it is unnecessary to confine that the number of
accessing users to be equal to the number of sensing users in the
joint optimization. However, if the number of sensing users is
independent, there will be one more integer optimization variable,
and thus the complexity of joint optimization will be increased
while the performance gain is limited.

In this section, we have derived the joint optimization schemes of
spectrum sensing and accessing, namely joint resource allocation
algorithms by maximizing the average sum transmission rate. In fact,
according to the aforementioned relationship of time and feedback
resources, we can realize the tradeoff between them for a given
performance requirement. For example, we can reduce the total
feedback amount by increasing sensing time, which is appealing to
feedback limited systems. Additionally, it is worth pointing out
that although the above schemes are derived based on the exponential
activity model of primary network, they are applicable for an
arbitrary activity model. As analyzed above, as long as the
stationary probabilities for the spectrum to be busy $P_B$ and idle
$P_I$ are given, the corresponding joint resource allocation
algorithm can be obtained.

\section{Simulation Results and Performance Analysis}
In order to verify the validity of our theoretical claims, we
present several simulations in different scenarios. The simulation
parameters are set according to Tab.\ref{Tab3}. In order to show the
advantages of the proposed joint optimization schemes explicitly, we
compare them with the traditional fixed resource allocation schemes,
which have fixed $\tau$, $B$, $L$ and $K_0$ when given $T$ and
$\iota$. Hereafter, we use Algorithm 1, Algorithm 2 and Traditional
Algorithm to denote the proposed algorithm 1, algorithm 2 and the
fixed resource allocation algorithm, respectively. Note that the
simulation results are obtained by averaging the sum rate over 10000
channel samples.

\newcommand{\tabincell}[2]{\begin{tabular}{@{}#1@{}}#2\end{tabular}}
\begin{table*}\centering
\caption{Simulation Parameter Table} \label{Tab3}
\begin{tabular*}{10.37cm}{|c|c|c|}\hline
Parameter & Description & Value\\
\hline $N_t$& Antenna Number & 4\\
\hline $T$ & Length of Time Slot & 10ms\\
\hline $W$ & Spectrum Bandwidth & 5KHz\\
\hline $\alpha$ & Transition Rate from Busy to Idle & 0.9\\
\hline $\beta$ & Transition Rate from Idle to Busy & 0.1\\
\hline $P_1$ & The Lower Bound on Detection Probability & 0.9\\
\hline $P_0$ & The Upper Bound on False-Alarm Probability & 0.1\\
\hline $\iota$ & Feedback Bits & 10, 20, 30, 40\\
\hline $\xi$ & Sensing SNR & -6dB, -4dB, -2dB, 0dB\\
\hline $B$ & Codebook Size & Optimization Variable\\
\hline $L$ & Number of Accessing User & Optimization Variable\\
\hline $\tau$ & Sensing Duration & Optimization Variable\\
\hline
\end{tabular*}
\end{table*}

\begin{figure}[h] \centering
\includegraphics [width=0.5\textwidth] {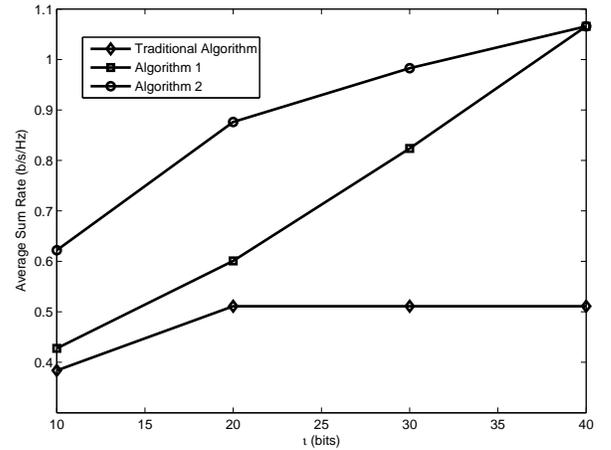}
\caption {Comparison of average sum rate with different feedback
constraints.} \label{Fig2}
\end{figure}

First, we investigate the impact of feedback constraint
on the joint optimization of spectrum sensing and accessing when
given $\xi=0$dB and $\varrho=15$dB. Tab.\ref{Tab1} shows the
resource allocation results with different feedback constraints.
Note that traditional algorithm allocates time or feedback source
separately as long as the requirements of detection and false-alarm
probabilities are met. It is found that, for the two proposed joint
optimization algorithms, when there is a strict feedback constraint,
i.e. small $\iota$, more time is allocated for spectrum sensing to
increase the precision of cooperative sensing. With the increase of
$\iota$, sensing time $\tau$ decreases accordingly while the number
of sensing users $L$ increases. This is because increasing $L$ is
more beneficial to improve sensing precision than increasing $\tau$.
Meanwhile, more time can be allocated to improve the average sum
rate. In addition, more feedback bits are distributed to CSI
conveyance, namely enlarging codebook size. There are two reasons:
first, when satisfying sensing precision, a large $B$ can decrease
interuser interference, and thus improve average sum rate; second,
limited by the probability of spectrum idle, namely $P_I$, further
increase in $L$ hardly adds the probability of spectrum accessing.
As a simple example, for algorithm 1, although $L$ is not limited by
$N_t$, it is still a relatively small value when $\iota$ is large.
Fig.\ref{Fig2} presents the corresponding average sum rates of the
above resource allocation results. It can be seen that the average
sum rate of traditional algorithm nearly keeps constant when $\iota$
is greater than 20, since the added feedback bits are used to
increase $L$. However, large $L$ is useless as discussed above. The
proposed joint optimization algorithms perform better than the
traditional scheme, and the performance gain becomes larger with the
increase of $\iota$. Algorithm 2 has an evident advantage over
algorithm 1 under the conditions of small $\iota$, because Algorithm
2 admits a small number of accessing users, so that the interuser
interference is decreased and codebook size $B$ can be enlarged.
With the increase of $\iota$, the performance gap between algorithm
2 and algorithm 1 reduces gradually until they have the same
performance, due to the same allocation results.

\begin{table}\centering
\caption{Joint optimization with different feedback constraints}
\label{Tab1}
\begin{tabular}{|c||c|c|c|c|c|}\hline
 & $\iota$ & 10 & 20 & 30 & 40 \\
\hline\hline & $B$ & 2 & 4 & 4 & 4
\\\cline{2-6} $\textmd{Traditional Algorithm}$& $L$ & 2 & 4 & 12 & 24 \\
\cline{2-6} & $\tau$ & 2.0 & 2.0 & 2.0 & 2.0 \\\hline & $B$ & 2 & 4&
7 & 9
\\\cline{2-6} $\textmd{Algorithm 1}$& $L$ & 2 & 4 & 2 & 4 \\
\cline{2-6} & $\tau$ & 0.8452 & 0.4467 & 0.8361 & 0.4467 \\\hline &
$B$ & 4 & 9 & 9 & 9
\\\cline{2-6} $\textmd{Algorithm 2}$& $L$ & 2 & 2 & 3 & 4 \\
\cline{2-6} & $\tau$ & 0.8361 & 0.8361 & 0.5789 & 0.4467 \\\hline
\end{tabular}
\end{table}

\begin{figure}[h] \centering
\includegraphics [width=0.5\textwidth] {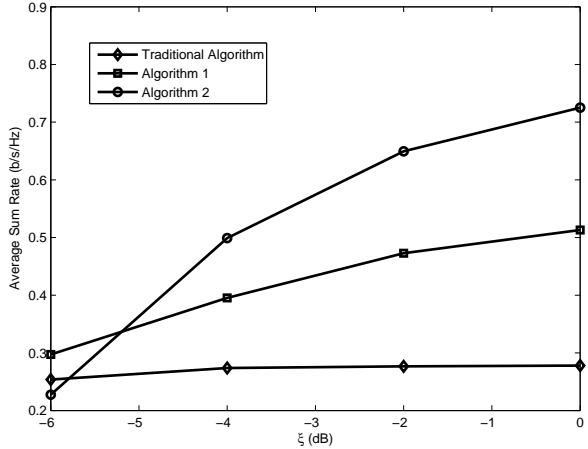}
\caption {Comparison of average sum rate with different sensing
SNRs.} \label{Fig3}
\end{figure}

Second, we study the role of sensing SNR $\xi$ in joint optimization
of spectrum sensing and accessing when given $\iota=15$ and
$\varrho=15$dB. As we know, sensing SNR has two contrary effects on
joint optimization and thus average sum rate. On one hand, high
$\xi$ can reduce the resource consumption for spectrum sensing,
hence more time and feedback can be used for accessing. On the other
hand, high $\xi$ means strong interference from primary network and
results in the descent of average sum rate. Herein, we consider the
scenario with low sensing SNR. As seen in Tab.\ref{Tab2}, when
sensing SNR is quite low, such as $-6$dB, feedback bits are
allocated to spectrum sensing as much as possible in order to
enhance the sensing precision through cooperation. For example,
algorithm 2 uses the upper bound of the feedback bits for sensing.
Once sensing SNR increases, more bits are used for CSI feedback,
which can achieve more performance gains as analyzed above. With the
further increase of sensing SNR, the allocation of feedback resource
is unchanged, while more and more time is used for spectrum
accessing. Examining the resultant average sum rate in
Fig.\ref{Fig3}, it is found that algorithm 2 performs even worse
than traditional algorithm when $\xi$ is equal to -6dB, this is
because feedback resource is under-utilized based on the allocation
scheme in this case. Specifically, the total number of used feedback
bits is $12$, but there are 3 bits left that are not enough to
enlarge codebook size and are not allowed to used for sensing
because the upper of the number of sensing bits is approached. As
sensing SNR increases, algorithm 2 outperforms the other two
algorithms quickly. More interestingly, algorithm 2 achieves the
performance advantage even with small number of feedback bits, which
shows its high feedback utilization efficiency.

\begin{table}\centering
\caption{Joint optimization with different sensing SNRs}
\label{Tab2}
\begin{tabular}{|c||c|c|c|c|c|}\hline
 & $\xi$ & -6dB & -4dB & -2dB & 0dB \\
\hline\hline & $B$ & 3 & 3 & 3 & 3
\\\cline{2-6} $\textmd{Traditional Algorithm}$& $L$ & 3 & 3 & 3 & 3 \\
\cline{2-6} & $\tau$ & 5.0 & 5.0 & 5.0 & 5.0 \\\hline & $B$ & 2 & 3&
3 & 3
\\\cline{2-6} $\textmd{Algorithm 1}$& $L$ & 7 & 3 & 3 & 3 \\
\cline{2-6} & $\tau$ & 3.3875 & 2.4268 & 1.0250 & 0.5883 \\\hline &
$B$ & 2 & 6 & 6 & 6
\\\cline{2-6} $\textmd{Algorithm 2}$& $L$ & 4 & 2 & 2 & 2 \\
\cline{2-6} & $\tau$ & 4.9347 & 3.3804 & 1.4812 & 0.8361 \\\hline
\end{tabular}
\end{table}

\section{Conclusions}
A major contribution of this paper is the construction of a joint
optimization framework of spectrum sensing and accessing in a
multiuser MISO cognitive network. Based on this framework, we
present two algorithms to jointly allocate time and feedback
resources, namely determining the duration of sensing time, the
number of sensing users, the number of accessing users, and CSI
quantization codebook size, by maximizing the average sum
transmission rate. When given a performance requirement, this
framework reveals the intrinsic relationship of these two resources.
As a result, the resource transformation can be realized according
to the characteristics of the considered network.

\appendix
For limited feedback ZFBF, since cognitive BS only has partial CSI
of the accessing users, there is still residual interuser
interference. In order to achieve the pdf of SINR, it is necessary
to reveal the relationship between codebook size and the residual
interference. According to the theory of random vector quantization
\cite{RVQ}, the relationship between the original and the quantized
channel direction vectors is given by
\begin{equation}
\tilde{\textbf{h}}_k=\sqrt{1-a}\hat{\textbf{h}}_k+\sqrt{a}\textbf{s},\label{app1}
\end{equation}
where $\hat{\textbf{h}}_k$ is the optimal quantization codeword
based on the codeword selection criteria (\ref{eqn30}).
$a=\sin^2\left(\angle\left(\tilde{\textbf{h}}_k,\hat{\textbf{h}}_k\right)\right)$
is the magnitude of the quantization error, and $\textbf{s}$ is a
unit norm vector isotropically distributed in the nullspace of
$\hat{\textbf{h}}_k$, and is independent of $a$. Therefore, the
interference term from the $u$th user to the $k$th user can be
expressed as
\begin{eqnarray}
|\textbf{h}_k^H\textbf{w}_{u}|^2&=&\|\textbf{h}_k\|^2|\tilde{\textbf{h}}_k^H\textbf{w}_{u}|^2\nonumber\\
&=&\|\textbf{h}_k\|^2
\bigg((1-a)|\hat{\textbf{h}}_k^H\textbf{w}_{u}|^2+a|\textbf{s}^H\textbf{w}_{u}|^2\nonumber\\
&&+2\sqrt{a(1-a)}|\textbf{w}_{u}^H\textbf{s}\hat{\textbf{h}}_k^H\textbf{w}_{u}|\bigg)\nonumber\\
&=&a\|\textbf{h}_k\|^2|\textbf{s}^H\textbf{w}_{u}|^2,\label{app2}
\end{eqnarray}
where (\ref{app2}) follows from the fact that both $\textbf{w}_{u}$
and $\textbf{s}$ are in the nullspace of $\hat{\textbf{h}}_k$,
namely $\hat{\textbf{h}}_k^H\textbf{w}_{u}=0$ and
$\textbf{s}\hat{\textbf{h}}_k^H=\textbf{0}$. Substituting
(\ref{app2}) into (\ref{eqn12}), we have
\begin{eqnarray}
\rho_k&=&\frac{|\textbf{h}_{k}^H\textbf{w}_{k}|^2}{1/\gamma+a\|\textbf{h}_{k}\|^2\sum\limits_{u=1,u\neq
k}^{N_t}|\textbf{s}^H\textbf{w}_{u}|^2}\nonumber\\
&\stackrel{d}{=}&\frac{\chi_2^2}{1/\gamma+a\|\textbf{h}_{k}\|^2\sum\limits_{u=1,u\neq
k}^{N_t}\beta(1,N_t-2)}\label{app3}\\
&\stackrel{d}{=}&\frac{\chi_2^2}{1/\gamma+\sum\limits_{u=1,u\neq
k}^{N_t}\Gamma(N_t-1,\delta)\beta(1,N_t-2)}\label{app4}\\
&\stackrel{d}{=}&\frac{\chi_2^2}{1/\gamma+\delta\chi_{2(N_t-1)}^2},\label{app5}
\end{eqnarray}
where $\delta=2^{-\frac{B}{N_t-1}}$ and $\stackrel{d}{=}$ denotes
the equality in distribution. $\beta(x,y)$ represents the Beta
distribution, whose probability density function is given by
$g(t)=\frac{t^{x-1}(1-t)^{y-1}}{\textmd{B}(x,y)}$, where
$\textmd{B}(x,y)=\frac{(x-1)!(y-1)!}{(x+y-1)!}$ is the Beta
function. (\ref{app3}) follows from the facts that $\textbf{w}_{k}$
of unit norm is independent of $\textbf{h}_{k}^H\textbf{w}_{k}|^2$,
so $\textbf{h}_{k}^H\textbf{w}_{k}$ is a complex Gaussian
distributed random variable with zero mean and unit variance. Then
$|\textbf{h}_{k}^H\textbf{w}_{k}|^2$ is $\chi_2^2$ distributed. In
addition, $|\textbf{s}^H\textbf{w}_{u}|^2$ is $\beta(1,N_t-2)$,
because $\textbf{s}$ and $\textbf{w}_{u}$ are i.i.d. isotropic
vectors in the $N_t-1$ dimensional null space of
$\hat{\textbf{h}}_k$ \cite{RVQ}. (\ref{app4}) is derived since
$a\|\textbf{h}_{k}\|^2$ is $\Gamma(N_t-1,\delta)$ distributed
according to the theory of quantization cell approximation
\cite{QCA}. Moreover, (\ref{app5}) holds true since the product of a
$\Gamma(N_t-1,\delta)$ distributed random variable and a
$\beta(1,N_t-2)$ distributed random variable is $\delta\chi_2^2$
distributed \cite{Modeselection}. Note that the sum of $N_t-1$
independent $\chi_2^2$ distributed random variables is
$\chi_{2(N_t-1)}^2$ distributed. Let $y\sim\chi_{2(N_t-1)}^2$ and
$z\sim\chi_2^2$, we can derive the cdf and pdf of $\rho_k$ as
follows
\begin{eqnarray}
F_{\rho_k}(x)&=&P\left(\frac{z}{1/\gamma+\delta y}\leq x\right)\nonumber\\
&=&\int_0^{\infty}F_{Z|Y}\big(x(1/\gamma+\delta y)\big)f_Y(y)dy\nonumber\\
&=&\int_0^{\infty}\big(1-\exp(-x(1/\gamma+\delta
y))\big)\nonumber\\
&&\times\frac{y^{N_t-2}}{\Gamma(N_t-1)}\exp(-y)dy\nonumber\\
&=&1-\frac{\exp\left(-x/\gamma\right)}{\left(1+\delta
x\right)^{N_t-1}},\label{app6}
\end{eqnarray}
and
\begin{eqnarray}
f_{\rho_k}(x)&=&F_{\rho_k}^{'}(x)\nonumber\\
&=&1/\gamma\exp\left(-x/\gamma\right)\left(1+\delta
x\right)^{-(N_t-1)}\nonumber\\
&&+\delta(N_t-1)\exp\left(-x/\gamma\right)\left(1+\delta
x\right)^{-N_t},\label{app7}
\end{eqnarray}
respectively, where $F_{Z|Y}(\cdot)$ is the conditional cdf of $z$
when given $y$, $f_Y(\cdot)$ is the pdf of $y$, $\Gamma(\cdot)$ is
the Gamma function. (\ref{app6}) holds true because
$\frac{\exp(-(1+\delta x)y)((1+\delta x)y)^{N_t-2}}{\Gamma(N_t-1)}$
is the pdf of $(1+\delta x)y$.

\end{document}